\documentclass[%
 reprint,
 superscriptaddress,
 amsmath,amssymb,
 aps,
]{revtex4-2}

\usepackage{graphicx}
\usepackage{float}
\usepackage{dcolumn}
\usepackage{bm}
\usepackage{siunitx}
\usepackage{xcolor}
\usepackage{booktabs}
\usepackage{natbib, hyperref}
\hypersetup{
	colorlinks=true,
	linkcolor=blue,      
	urlcolor=blue,
	citecolor=blue
}

\begin{document}

\preprint{APS/123-QED}

\title{Extreme Ultraviolet High-Harmonic Interferometry of Excitation-Induced Bandgap Dynamics in Solids}

\author{Lisa-Marie Koll,$^{1,\dagger}$ Simon Vendelbo Bylling Jensen,$^{2,\dagger}$ Pieter J. van Essen,$^{3, \dagger}$ Brian de Keijzer,$^{3, \dagger}$ Emilia Olsson,$^{3}$ Jon Cottom,$^{3}$ Tobias Witting,$^{1}$ Anton Husakou,$^{1}$ Marc J. J. Vrakking,$^{1}$ Lars Bojer Madsen,$^{2}$ Peter M. Kraus,$^{3,4}$ and Peter J{\"u}rgens,$^{1,3,*}$}

\address{$^{}$Max-Born-Institute for Nonlinear Optics and Short Pulse Spectroscopy, Max-Born-Str. 2A, 12489 Berlin, Germany\\
$^{2}$Department of Physics and Astronomy, Ny Munkegade 120, Aarhus University, DK-8000, Denmark\\
$^{3}$Advanced Research Center for Nanolithography (ARCNL), Science Park 106, 1098 XG Amsterdam, Netherlands\\
$^{4}$Department of Physics and Astronomy, and LaserLaB, Vrije Universiteit, De Boelelaan 1105, 1081 HV Amsterdam, Netherlands \\
$^{\dagger}$ These authors contributed equally: L.-M. Koll, S. V. B. Jensen, P. J. v. Essen, B. d. Keijzer}

\email{$^{*}$juergens@mbi-berlin.de}


\begin{abstract} 

Interferometry is a fundamental technique in physics, enabling precise measurements through the interference of waves. High-harmonic generation (HHG) in solids has emerged as a powerful method for probing ultrafast electronic dynamics within crystalline structures.

In this study, we employed extreme ultraviolet (XUV) high-harmonic interferometry with phase-locked XUV pulse pairs to investigate excitation-induced bandgap dynamics in solids. Our experiments on amorphous SiO$_2$ and crystalline MgO, complemented by analytical modeling and semiconductor Bloch equation simulations, reveal a correlation between transient bandgap modifications and variations in the phase of harmonic emission. These findings suggest a potential pathway for sub-cycle, all-optical control of band structure modifications, advancing prospects for petahertz-scale electronic applications and attosecond diagnostics of carrier dynamics.

\end{abstract}

\maketitle

\section{Introduction}
%
Optical interferometry stands as a cornerstone in the annals of optical sciences, tracing its roots back to seminal works such as the observation of the diffraction pattern in single and double-slit experiments. In the field of ultrafast physics, interferometry has been widely used for the characterization of ultrashort laser pulses using the technique of Spectral Phase Interferometry for Direct Electric-field Reconstruction  (SPIDER) \cite{Iaconis_1999}. Further, the transient alteration of the optical properties of transparent solids has been studied by spectral interferometry yielding information on the dynamic modification of the refractive index during ultrafast light-matter interaction \cite{Froehly_1973, Lepetit_1995}. 
The advent of attosecond science, marked by milestone achievements such as the experimental generation of attosecond pulse trains \cite{Huillier_1993, Paul_2001} and isolated attosecond pulses \cite{Hentschel_2001} – recognized with the Nobel Prize in Physics in 2023 \cite{lHuillier_2024, Krausz_2024, Agostini_2024} – has profoundly impacted the field of ultrafast and nonlinear optics \cite{Corkum_2007}. It has not only facilitated the characterization of attosecond pulses \cite{Tzallas_2003} but has also enabled the probing of ultrafast electron wavefunction dynamics \cite{Azoury_2019} and field-induced tunneling ionization \cite{Kneller_2022}. \\
With the pioneering demonstration of high-order harmonic generation (HHG) in solids \cite{Chin_2001, Ghimire_2011}, HHG in solids has emerged as a powerful tool for probing ultrafast electronic dynamics \cite{Nie_2023, van_2023} and band structure properties in condensed matter systems \cite{Lanin_2017, Luu_2015, Schubert_2014, Zaks_2012}.  
In parallel, the concepts of solid-state HHG and interferometry have first been merged to track the intensity-dependence of the dipole phase \cite{Lu_2019}. For atomic HHG, the dipole phase can be viewed as the phase accumulated by an electron wavepacket during its excursion, semiclassically described by the three-step model \cite{Corkum_1993, Lewenstein_1994, Lewenstein_1995}. Lu et al. \cite{Lu_2019} investigated the spatial interference pattern in the far field, generated by two phase-locked NIR pulses focused to two separate spots in a solid sample. Their results identified intensity-dependent dipole phase variations that were attributed to changes in the nonlinear polarization. In a more recent study, Uchida et al. employed a Mach-Zehnder interferometer to track the real-time dynamics of Floquet states in WSe$_2$\cite{Uchida_2024}.\\
Our work builds upon these studies by employing spectral interferometry of phase-locked extreme ultraviolet (XUV) pulse pairs to investigate excitation-induced bandgap dynamics in bulk solids. We focus on amorphous SiO$_2$ and crystalline MgO samples, both of which have been subjects of previous HHG studies \cite{Juergens_2020, You_2017sio2, Roscam_2022, You_2017, Juergens_2024}. In contrast to the earlier experiments performed by Lu et al., we investigate the spectral interference generated by two collinearly-propagating phase-locked pulses. Hence, rather than examining intensity-dependent harmonic phase changes, we investigate excitation-induced phase variations between the two generated XUV pulses.
Our experimental results are corroborated by numerical calculations based on an analytical model alongside solution of the semiconductor Bloch equations (SBEs) \cite{Golde2006, Haug2009}. We observe high-harmonic phase shifts of opposite sign in amorphous SiO$_2$ and crystalline MgO samples and correlate this observation with transient, excitation-induced alterations of the electronic structure. As a result, our work introduces a potential technique for all-optical control and characterization of band structure modifications on sub-cycle timescales. These advancements hold promise for transformative applications in attosecond carrier dynamics diagnostics.

\section{Experimental Setup}
\begin{figure}[ht]
    \centering\includegraphics[width=\linewidth]{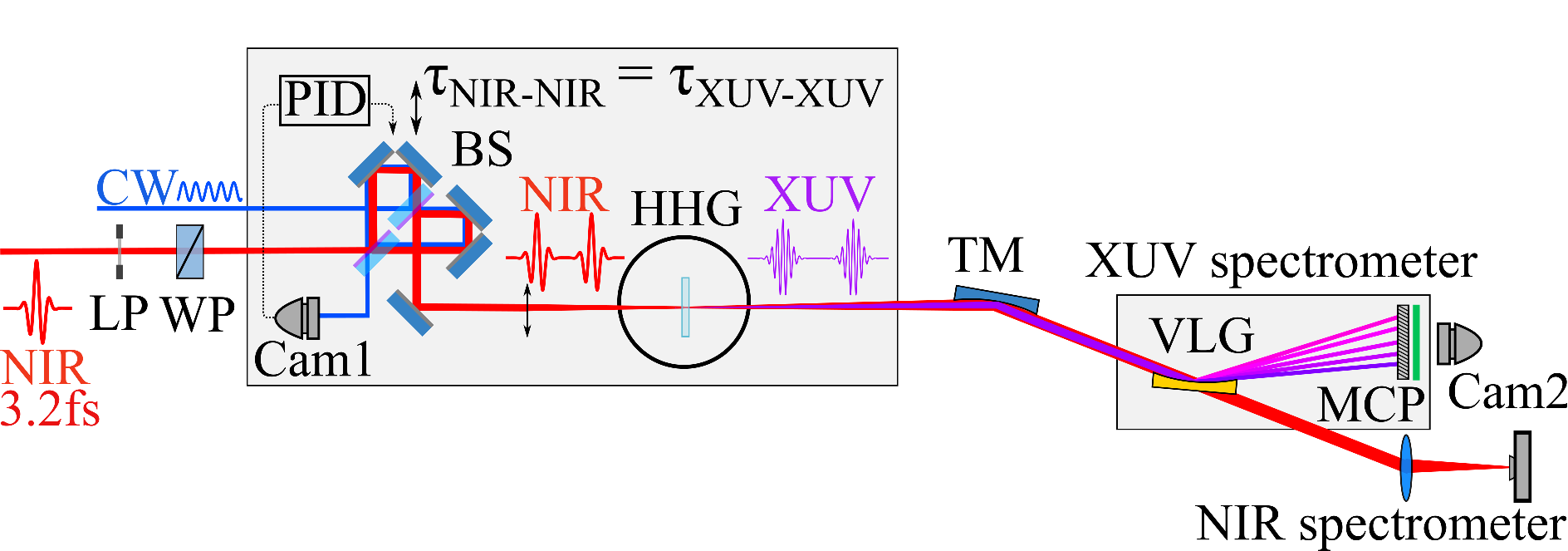}
    \caption{Experimental Setup. A near-infrared (NIR), few-cycle laser pulse was split into two identical copies and focused into a bulk solid sample. The resulting phase-locked extreme ultraviolet (XUV) pulse pair was analyzed with the help of an XUV spectrometer, while the fundamental NIR beam was analyzed with a near-infrared spectrometer. LP: Half-wave plate, WGP: Wire-grid polarizer, CW: Continuous wave stabilization laser, PID: Proportional–integral–differential controller, Cam 1: CMOS camera, BS: Beam splitters, TM: Toroidal mirror, VLG: Variable line-spacing grating, MCP: Micro-channel plate + phosphor screen, Cam2: CCD camera.}
    \label{fig:setup}
\end{figure}

In our experiments, we utilized a few-cycle, near-infrared (NIR) waveform with a central wavelength of \SI{750}{\nano\meter} and a full-width half-maximum intensity pulse duration of approximately \SI{4}{\femto\second} (measured with a home-built SEA-F-SPIDER \cite{Witting_2011}). This NIR waveform was divided into two identical copies using a passively and actively stabilized Mach-Zehnder interferometer (MZI), as described in Ref.~\cite{Koll_2022}.
At the output of the MZI, the two identical, phase-locked NIR pulses propagated collinearly. The beams were focused onto a bulk solid sample by a single spherical mirror with a focal length of \SI{750}{\milli\meter}. Control over the total pulse energy entering the MZI was achieved using a combination of an achromatic zero-order half-wave plate and a broadband wire-grid polarizer.
The temporal separation between the two phase-locked NIR pulses (referred to as $\tau_{\rm{NIR-NIR}}$) was adjusted by a computer-controlled delay stage within the MZI. Additionally, a second piezo stage was used to actively stabilize the MZI using the interferometric signal of a co-propagating continuous-wave laser as a reference.
Depending on the thickness of the solid samples used, we optimized the pulse duration using a pair of fused silica wedges, allowing for precise control of the dispersion. This optimization ensured optimal contrast in the interferogram and well-separated harmonics, facilitating the analysis of the phase for individual harmonics. The wedge position was kept constant throughout all measurements to ensure consistency and reproducibility.
The intense NIR pulses interacted with the solid sample, leading to the generation of HHG extending into the XUV spectral range. Due to the phase-locking between the two identical NIR pulses, a phase-locked XUV pulse pair (separated by $\tau_{\rm{XUV-XUV}} = \tau_{\rm{NIR-NIR}}$) was produced. The XUV radiation was recollimated and directed towards an XUV spectrometer with the help of a toroidal mirror (TM in Fig.~\ref{fig:setup}).
To spectrally resolve the individual harmonics, the beams were reflected by a variable line-spacing grating (labeled VLG in Fig.~\ref{fig:setup}) before being detected by a double-stack micro-channel plate (MCP) and a phosphor screen monitored with a CCD-camera from outside the vacuum chamber. The VLG could be removed from the beam path with the help of a motorized translation stage to enable the analysis of the NIR beams using a commercial fiber spectrometer (Ocean Optics FLAME-S-VIS-NIR, NIR spectrometer in Fig.~\ref{fig:setup}).
XUV and NIR spectra were acquired as a function of the NIR intensity at various NIR-NIR delays. Throughout the measurements, the integrity of the samples was ensured by monitoring the stability of the HHG signal and potential scattering of the fundamental beam from permanent, laser-induced modifications.
As demonstrated in prior studies \cite{Luu_2015}, the combination of thin samples and few-cycle laser pulses ensured that the solid samples could withstand extremely high field strengths on the order of several \si{\volt \per \angstrom}.


\section{Experimental Results}
\begin{figure}[ht]
    \centering\includegraphics[width=\linewidth]{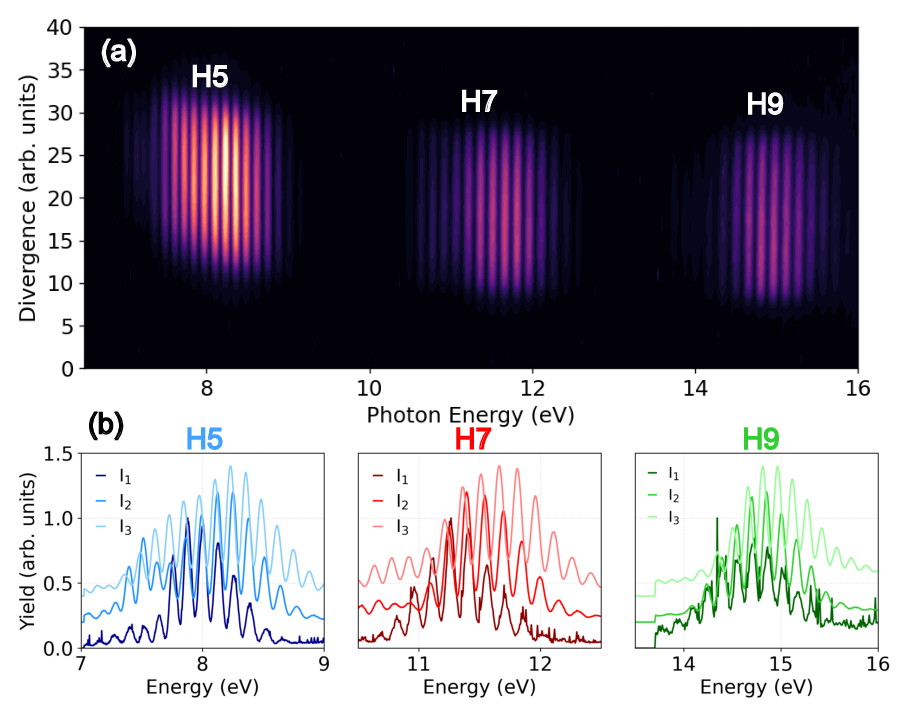}
    \caption{XUV interferometry in bulk SiO$_2$. (a) Interferometric pattern of harmonics 5,7 and 9 at an NIR-NIR delay of \SI{30}{\femto\second}. (b) Spatially integrated harmonic signal of the individual odd harmonics obtained at different NIR peak intensities ($I_1~=$~\SI{15.7}{\tera\watt\per\centi\meter\squared}, $I_2~=$~\SI{32.3}{\tera\watt\per\centi\meter\squared}, $I_3~=$~\SI{46.7}{\tera\watt\per\centi\meter\squared}) projected onto the energy axis. The spectra are offset vertically (by 0.2 for $I_2$ and 0.4 for $I_3$) for visibility.}
    \label{fig:fig_1}
\end{figure}

\begin{figure}[ht]
    \centering\includegraphics[width=\linewidth]{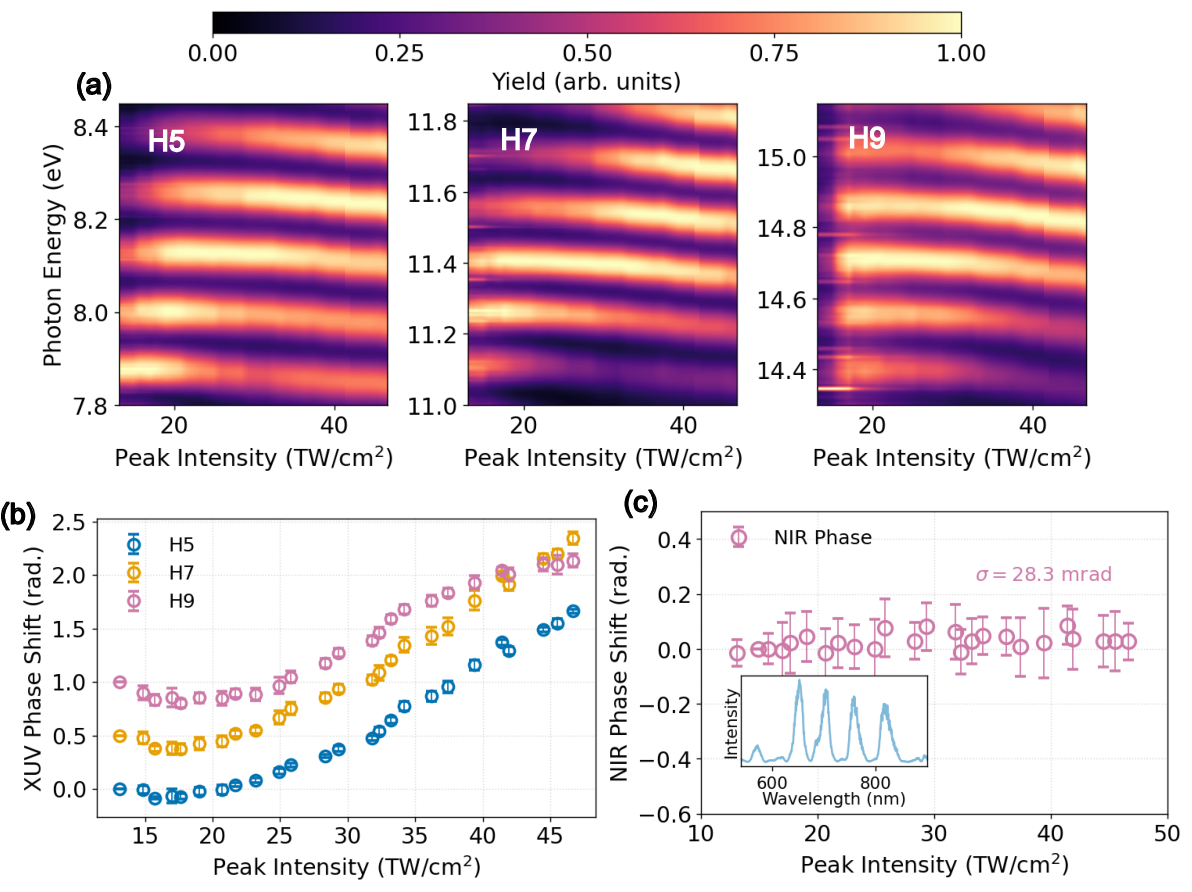}
    \caption{Extraction of the $\overline{\Delta \phi}_{\rm{XUV}}$ and $\overline{\Delta \phi}_{\rm{NIR}}$ in SiO$_2$. (a) 2D maps of the intensity dependence of the spectral interferometry signal of the observed high harmonics. (b) Extracted XUV phase shift as a function of the NIR peak intensity (Phases of H7 and H9 are offset vertically by \SI{0.5}{\radian} and \SI{1}{\radian} for visibility). (c) NIR spectral interferometry in SiO$_2$. The pink circles illustrate the intensity dependence of the NIR phase shift $\overline{\Delta \phi}_{\rm{NIR}}$. The blue line in the inset shows the NIR-interference pattern.} 
    \label{fig:fig_2}
\end{figure}

\begin{figure}[ht]
    \centering\includegraphics[width=\linewidth]{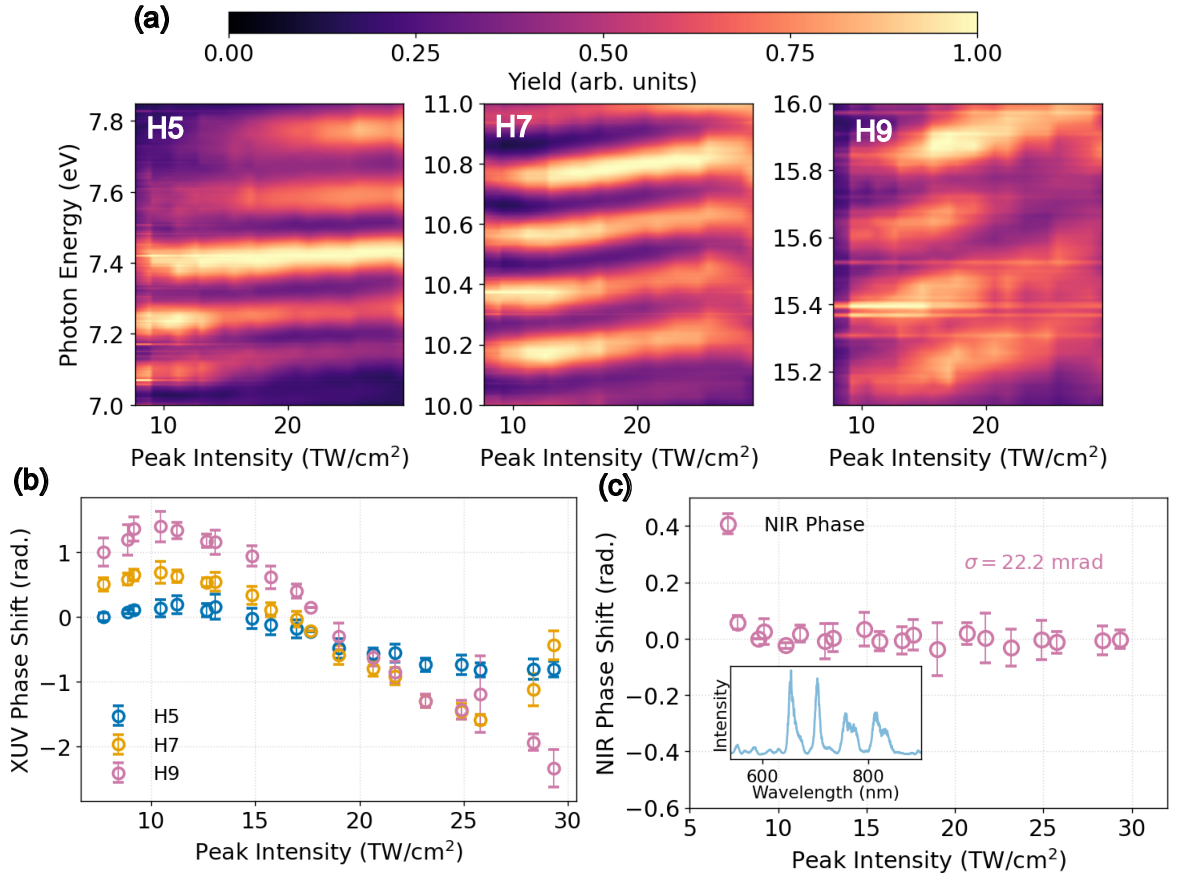}
    \caption{Extraction of the $\overline{\Delta \phi}_{\rm{XUV}}$ and $\overline{\Delta \phi}_{\rm{NIR}}$ in MgO. (a) 2D maps of the intensity dependence of the spectral interferometry signal of the observed high harmonics. (b) Extracted XUV phase shift as a function of the NIR peak intensity (Phases of H7 and H9 are offset vertically by \SI{0.5}{\radian} and \SI{1}{\radian} for visibility). (c) NIR spectral interferometry in MgO. The pink circles illustrate the intensity dependence of the NIR phase shift $\overline{\Delta \phi}_{\rm{NIR}}$. The blue line in the inset shows the NIR-interference pattern.}
    \label{fig:fig_3}
\end{figure}

In our experiments, we investigated the XUV spectral interferometry signal obtained from amorphous SiO$_2$ (UV-grade Fused Silica, Corning7980, \SI{10}{\micro\meter} thickness) and crystalline MgO (SurfaceNet, \SI{100}{\micro\meter} thickness, measurements were conducted along the [100] crystallographic direction). For both materials, we observed odd harmonics of orders five to nine, reaching photon energies of approximately $\sim$\SI{16}{\electronvolt}. The maximum detected photon energy was determined by the geometry of the detection unit and does not represent the cutoff of the HHG spectrum. A representative high-harmonic interferogram obtained with an NIR-NIR delay of \SI{30}{\femto\second} in SiO$_2$ excited at a peak intensity of \SI{32.3}{\tera\watt\per\centi\meter\squared} is shown in Fig.~\ref{fig:fig_1}$\,$(a). A well-defined interference pattern, consisting of periodic minima and maxima, is visible in all detected harmonics (labeled H5, H7, and H9).
Despite the substantial bandwidth of the driving NIR pulses, which support durations close to a single optical cycle, we observed well-separated odd harmonics. This separation results from the positive chirp of the NIR pulses, leading to constructive interference of attosecond bursts emitted during each half-cycle \cite{Chang_1998}. Ideally, the spacing between interference fringes remains constant across the high-harmonic spectrum; however, slight deviations among individual harmonics can be attributed to detector calibration inaccuracies and potential dispersion effects within the solid samples. \\
Before proceeding with additional analysis of the interferometric measurements, it is essential to ensure the setup's stability. A comprehensive characterization and analysis of the stability assessment are provided in the Supporting Information as well as in Ref.~\cite{Koll_2022} demonstrating a stability of below \SI{10}{\atto\second}. \\
%
To investigate the field-intensity impact on the interferometric fringe pattern, we modulated the peak intensity of the two phase-locked NIR pulses while maintaining a constant splitting ratio of $50:50$ in the MZI. By progressively increasing the intensity of both pulses, we aimed to assess how the interaction of the first NIR pulse, $E_{1}(\omega, t)$, with the sample influences the subsequent interaction between the second NIR pulse, $E_{2}(\omega, t-\tau)$, and the sample. This influence manifests as a shift in the minima and maxima observed in the interferometric spectra, reflecting a phase shift between the two generated phase-locked XUV fields, termed $\Delta \phi_{\rm{XUV}}$. Figure~\ref{fig:fig_1}$\,$(b) illustrates such shifts in the interferometry signal obtained at different NIR peak intensities. Generally, the minima and maxima shift towards lower energies as the NIR peak intensity increases. 
The collective shift of the spectral interference fringes, associated with $\Delta \phi_{\rm{XUV}}$, as a function of the NIR intensity is summarized in Fig.~\ref{fig:fig_2}$\,$(a). It is apparent that at moderate NIR peak intensities of up to $\sim \,$\SI{25}{\tera\watt\per\centi\meter\squared}, the location of the minima and maxima in the interference pattern remains almost constant, while at peak intensities above \SI{30}{\tera\watt\per\centi\meter\squared}, a redshift of the fringe pattern becomes similarly visible in the interferograms of all observed harmonics. In addition to the fringes moving towards lower photon energy as the intensity increases, the energy of the harmonics moves up. We emphasize that this blueshift of the individual harmonics does not affect the determination of $\Delta \phi_{\rm{XUV}}$. \\
To quantify the high-harmonic phase shift, we employed the Takeda algorithm \cite{Takeda_1982}. This algorithm extracts the spectral phase from the measured fringe patterns by filtering out the AC component in the Fourier domain. We compared all retrieved phases to those obtained in the initial measurement conducted at the lowest NIR intensity. Consequently, the initial measurement serves as our phase reference ($\Delta \phi_{\rm{XUV}}=0$), and subsequent measurements reveal the relative phase shift compared to this baseline. It is important to note that this reference is arbitrary since the absolute phase is unknown; our interest lies solely in the relative phase — namely, how the phase $\Delta \phi_{\rm{XUV}}$ changes with varying intensity levels.
To quantify the phase shift $\Delta \phi_{\rm{XUV}}$ for a given harmonic order, we calculated a weighted average of the phase shift across the measured intensity distribution of an individual harmonic order. This approach involves using the measured intensity profile of each harmonic as a weighting factor, ensuring that any small phase shifts within a harmonic peak are incorporated into the overall variance of the mean phase shift value.
The weighted mean phase shift, $\overline{\Delta \phi}_{\rm{XUV}}$ of an individual harmonic order can be expressed as 
\begin{equation}
    \overline{\Delta \phi}_{\rm{XUV}} = \frac{\sum_{i=1}^{N} Y_i \Delta \phi_{i}}{\sum_{i=1}^{N} Y_i }
\end{equation}
where $Y_{i}$ represents the measured harmonic signal yield at frequency $i$, where $i$ refers to frequencies around the corresponding harmonic. The result of this analysis is shown in Fig.~\ref{fig:fig_2}$\,$(b), where the resulting phase shift is visualized as a function of the NIR peak intensity for the three observed harmonics. After an almost flat region (with the exception of a small minimum around \SI{17}{\tera\watt\per\centi\meter\squared}) up to $\sim$\SI{22}{\tera\watt\per\centi\meter\squared}, the extracted phase for all harmonics substantially increases by more than 1 rad before reaching its maximum value at the highest experimentally accessible (i.e., non-destructive) intensity. Unlike H5 and H7, the extracted phase of H9 shows a saturation behavior at the highest NIR intensities. \\
%
%
In interpreting the observed $\overline{\Delta \phi}_{\rm{XUV}}$, it is ambiguous whether the phase alteration arises from the XUV-generation process (i.e. HHG) or if it stems from a phase accumulation of the time-delayed, fundamental NIR pulse $E_{2}(\omega, t-\tau)$ propagating through a region within the sample potentially modified by $E_{1}(\omega, t)$. The interaction of$E_{1}(\omega, t)$ with the sample may induce modifications in the refractive index via the generation of an electron-hole plasma, leading to a phase accumulation $\Delta \phi_{\rm{NIR}}$ of the time-delayed second NIR pulse \cite{Juergens_2024_LPR}. Consequently, the resulting XUV phase would follow from the NIR phase via $\Delta \phi_{\rm{XUV}} = \mathcal{N} \Delta \phi_{\rm{NIR}}$, where $\mathcal{N}$ denotes the harmonic order.
To investigate the potential imprint of the NIR phase on the XUV phase, spectral interferometry measurements were conducted in the NIR spectral range. The measurements were conducted under the same conditions as the XUV interferometry experiments, utilizing identical samples and maintaining an identical beam path up to the detection stage (see Fig.~\ref{fig:setup}). Analogous to the procedure described for the XUV spectra, we used the Takeda algorithm to extract the intensity-dependent phase variation $\overline{\Delta \phi}_{\rm{NIR}}$ [purple circles in Fig.~\ref{fig:fig_2}(c)]. The observed NIR phase shows no significant intensity dependence with a standard deviation of only \SI{28.3}{\milli\radian} indicating that the observed change of $\Delta \phi_{\rm{XUV}}$ is not due to propagation effects and is not inherited from the NIR phase. 
The absence of a detectable electron-hole plasma signature in the NIR interferometry measurements can be attributed to the use of thin samples combined with short pulses, which together constrain both the interaction length and the resulting electron-hole density (a detailed explanation and numerical estimate are provided in the SI). \\
The same measurements as above were performed for mono-crystalline MgO samples. The results are summarized in Fig.~\ref{fig:fig_3}. As shown in Fig.~\ref{fig:fig_3}$\,$(a), the overall shift of the interferometric pattern points in the opposite direction. At high intensities above $\sim$\SI{15}{\tera\watt\per\centi\meter\squared}, a substantial shift of the interference pattern towards higher energies (blueshift) was observed. This behavior is further illustrated in Fig.~\ref{fig:fig_3}$\,$(b), where the extracted $\overline{\Delta \phi}_{\rm{XUV}}$ as a function of NIR peak intensity is shown for the observed harmonics. The NIR spectral interferometry results shown in Fig.~\ref{fig:fig_3}$\,$(c) again only display slight variations in the weighted mean NIR phase.
%
%

\begin{figure}[ht]
    \centering\includegraphics[width=\linewidth]{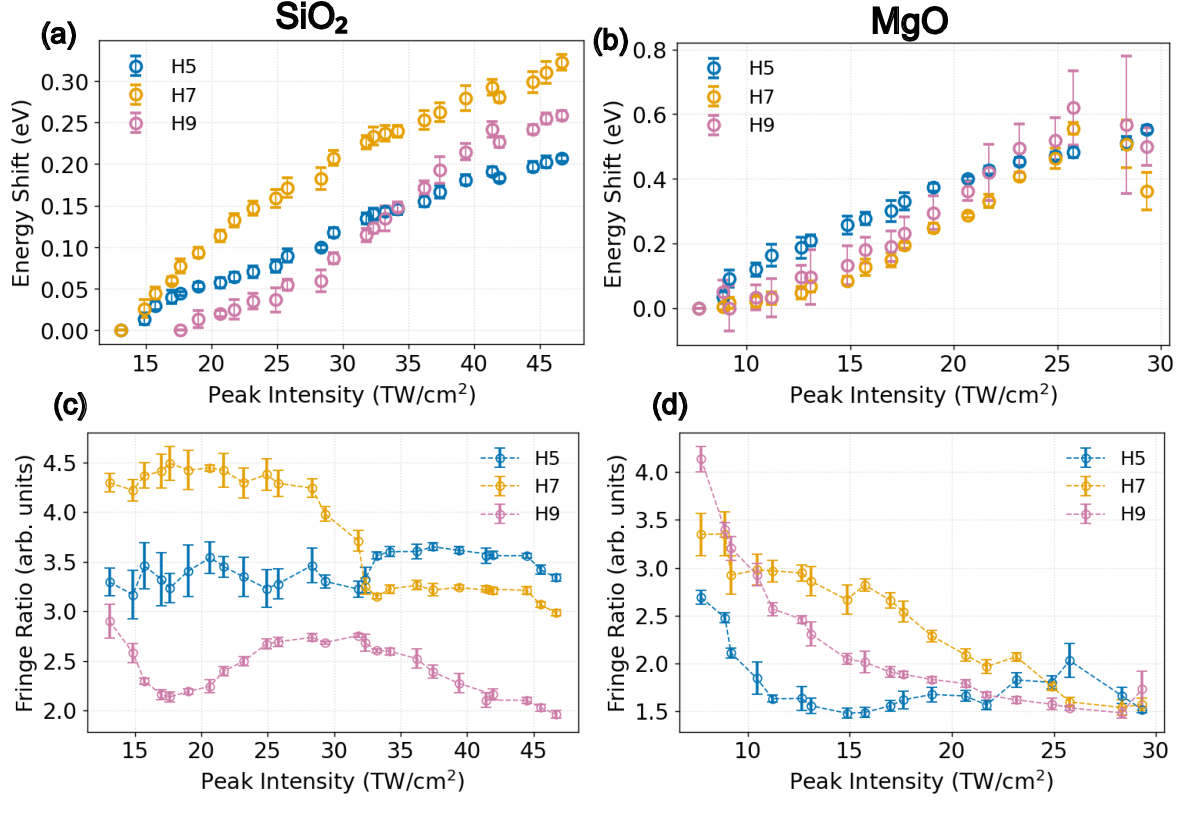}
    \caption{Extraction of energy shifts and fringe ratios from the XUV interferometry measurements. (a) Energy shift of the individual harmonics as a function of excitation intensity in SiO$_2$. (b) Same as in (a) for MgO. (c) Extracted fringe ratio of the observed odd harmonics as a function of the NIR peak intensity in SiO$_2$. (d) Same as in (c) but for MgO.}
    \label{fig:add_exps}
\end{figure}

Additional experimental observables extracted from the XUV interferometry measurements are summarized in Fig.~\ref{fig:add_exps}. Panels (a) and (b) illustrate the central energy shifts of the observed harmonics in SiO$_2$ and MgO, respectively, both exhibiting continuous blueshifts. Notably, the energy shifts in MgO are significantly larger, reaching up to \SI{0.6}{\electronvolt}, compared to a maximum of \SI{0.3}{\electronvolt} in SiO$_2$. Figures~\ref{fig:add_exps}(c) and (d) show the fringe contrast of the interference patterns observed in SiO$_2$ and MgO as a function of the NIR peak intensity. While SiO$_2$ shows no characteristic trend across the full range of applied NIR intensities, MgO exhibits a clear reduction in fringe contrast with increasing intensity across all observed harmonic orders.

\section{Modelling and Discussion}
\subsection{Modelling of XUV high-harmonic interferometry in solids}
\begin{figure}[ht]
    \centering\includegraphics[width=\linewidth]{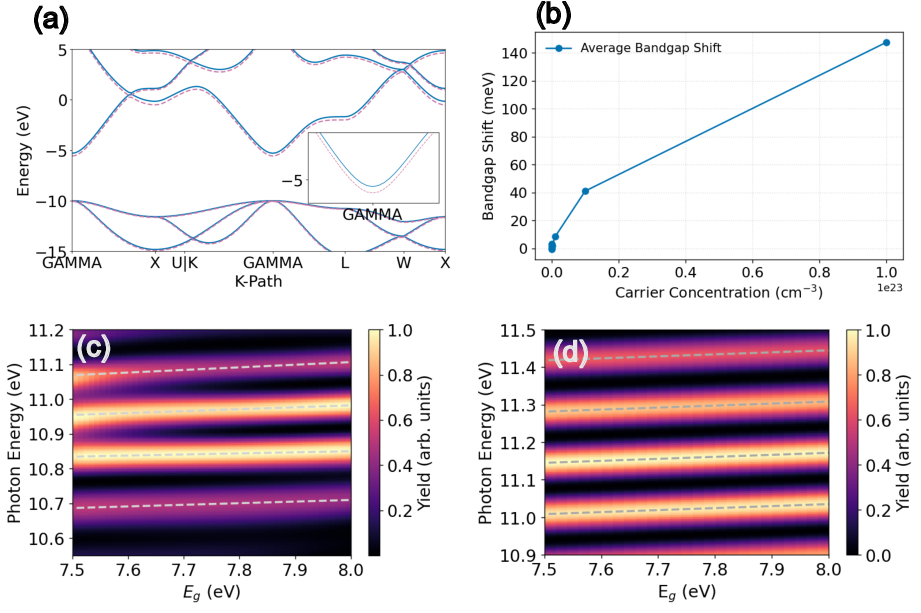}
    \caption{(a) DFT calculations of the band structure variations in MgO due to an excited carrier population. Purple lines show the bandstructure of the unperturbed crystalline system while the blue lines are obtained at an excited population of \SI{1e23}{\per\centi\meter}. (b) Averaged bandgap shift due to excitation of carriers. (c) Numerically simulated bandgap-dependent shift of the interference pattern of the seventh harmonic based on obtained by solving the SBEs. (d) Same as in (c) but obtained by the semi-classical analytical model.}
    \label{fig:theory}
\end{figure}

To support our experimental findings we performed analytical calculations and numerical simulations using a versatile set of models and tools. First, we computed the influence of an excited carrier population on the band structure of crystalline MgO using first-principle density-functional theory (DFT, details can be found in the SI). The comparison of ab-initio calculations of the electronic distribution in pristine and strongly-perturbed MgO reveals that due to state-blocking, the energy gap between the highest-lying valence band state and the lowest-lying conduction band state increases steadily as a function of the carrier concentration [see Fig.~\ref{fig:theory}(a) \& (b)]. At a carrier concentration of \SI{1e23}{\per\cubic\centi\meter} a bandgap widening of above \SI{140}{\milli\electronvolt} is predicted by our DFT calculations. \\
%

To model the spectral interferometry of HHG from solids we numerically solved the SBEs in a two-band tight-binding approximation for cubic MgO along the $\mathrm{\Gamma}$-X direction (for further details see SI) employing a phase-locked NIR pulse pair. Since excitation-induced variations in the electronic structure are not directly accounted for by the SBEs and are themselves extremely challenging to compute (see, e.g., Refs.~\cite{Tsaturyan_2022, Tsaturyan_2024}), we instead monitor the resulting high-harmonic interference pattern as a function of the bandgap $E_g$. This assumes that the conduction band population excited by the first strong NIR field dynamically modifies the bandgap, altering the HHG process induced by the second NIR field and leading to a variation in the high-harmonic phase.\\
For the simulations, we performed separate computations of single-pulse HHG for MgO with a bandgap varying from 7.5 to 8.0 eV in \SI{10}{\milli\electronvolt} steps. We subsequently compute the spectral interference of the emission with original bandgap at \SI{7.5}{\electronvolt} and the time-delayed emission with varied bandgap. In this approach, the bandgap change is treated as a parameter (but independently calculated in the aforementioned DFT simulations), which allows us to study how bandgap changes manifest themselves as shifts of spectral interference fringes.

Figure~\ref{fig:theory}(c) shows the bandgap-dependence of the seventh harmonic interference pattern obtained from an SBE simulation at a NIR peak intensity of \SI{30}{\tera\watt\per\centi\meter\squared} in MgO. A substantial frequency blueshift of the interference fringes for increasing bandgap (bandgap widening) is observed. The dashed gray lines in Fig.~\ref{fig:theory}(c) emphasize this blueshift, revealing a linear increase in the fringe frequency of approximately $0.05$.
These results resemble the experimental findings for MgO shown in Fig.~\ref{fig:fig_3}(a), suggesting a bandgap widening in the MgO crystals being responsible for the observed $\overline{\Delta \phi}_{\rm{XUV}}$. \\
To analytically link high-harmonic phase shifts and bandgap modifications, we have employed an additional analytical, semi-classical two-band model for the calculation of HHG in solids as introduced in \cite{deKeijzer_2024}. This model was numerically verified in \cite{deKeijzer_2024} by comparing its predicted phase shifts to the phase shifts obtained from numerical solutions of the two-band semiconductor Bloch equations in length gauge. In brief, in the analytical model, we calculated classical trajectories within a two-band system under the assumption of low carrier inversion. Excitation is restricted to the $\Gamma$-point, to ensure unambiguous separation of long and short trajectories which are not found using the SBEs, as these allow for excitation along the entire band structure \cite{Huttner_2016} according to the k-dependent transition dipole moment and band energy separation. \\
Electron wavepacket trajectories for different excitation times are determined by a purely classical integration of the group velocity $v_g^{\lambda}(k(t))$ over time:
\begin{equation}
    x^{\lambda}(t) = \int_{t_i}^{t} v_g^{\lambda}(k(\tau))d\tau ,
\end{equation}
with $\lambda \in \{e,h\}$. Recombination is assumed to occur when the spatial displacement between the electron and hole becomes zero:
\begin{equation}
    \Delta x (t_r) = x^e(t_r) - x^h(t_r) = 0. 
\end{equation}
Here, $t_r$ represents the recombination time. The photon energy emitted by a given trajectory corresponds to the energy difference $\Delta \varepsilon$ between the carriers at the moment of recombination. To assess the influence of bandgap variations on the phase of the associated interband current, we calculated the phase of the emitted light using the semi-classical action:
\begin{equation}
    S(t_r) = \int_{t_i}^{t_r} \Delta \varepsilon(k(\tau)) d\tau .
\end{equation}
From $S(t_r)$ we can evaluate the dipole phase \cite{Carlstroem2016} via
\begin{equation}
    \phi = \mathcal{N} \left(\omega_0 t_r + \frac{\pi}{2} \right) - S(t_r), 
\end{equation}
with $\mathcal{N}$ being the harmonic order. Under the assumption of small bandgap variations the resulting phase difference can be approximated by
\begin{equation}
    \label{eq:Pieters_expression}
    \Delta \phi \approx - \Delta E_g \Delta t \quad \text{with} \quad \Delta t = t_r - t_i .
\end{equation}
For MgO, we obtained a characteristic excursion time of $\Delta t \sim\,$\SI{1.5}{\femto\second} for the 7th harmonic order in \cite{deKeijzer_2024}, which was obtained by fitting the phase shifts from the 2-band SBEs [Fig.~\ref{fig:theory}(c)] with Eq.~\ref{eq:Pieters_expression}.
Results for the relationship between the frequency shift of the interference spectrum and the bandgap variation for the seventh harmonic obtained from the analytical model are presented in Fig.~\ref{fig:theory}(d). A linear relationship between the shift of the interference pattern and the bandgap variation $\Delta E_g$ with an average slope of $\sim0.05\,$ (indicated by the gray dashed lines, identical to the slope observed in the SBE results) can be observed.
Again, the qualitative agreement with the experimental results shown in Fig.~\ref{fig:fig_3} indicates that indeed photo-induced bandgap dynamics and associated variations of the dipole phase can be held responsible for the experimentally observed $\Delta \phi_{\rm{XUV}}$. \\
\begin{figure}[ht]
    \centering\includegraphics[width=\linewidth]{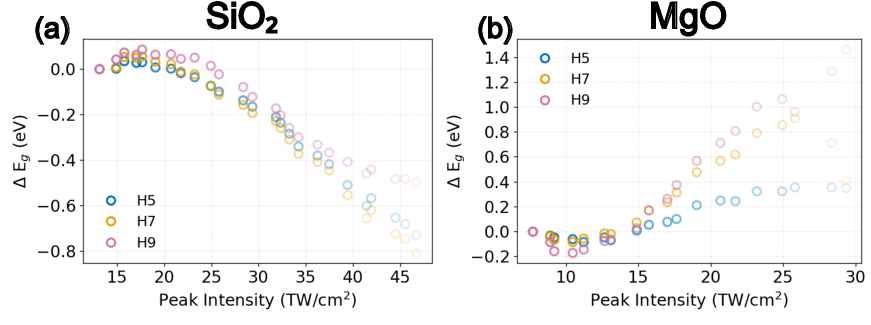}
    \caption{(a) Extracted bandgap variation as a function of the NIR peak intensity in SiO$_2$ obtained with the help of Eq.~\ref{eq:Pieters_expression}. (b) Same as in (a) but for MgO. Results in the high-intensity regime are faded due to the limited validity of Eq.~\ref{eq:Pieters_expression} in this regime and the underlying approximations.}
    \label{fig:final_figure}
\end{figure}
\subsection{Discussion and Interpretation of the results}
Since the photon energies of all observed high-order harmonics lie well above the bandgap of the used samples, we expect the interband recollisions to provide the strongest contribution to the harmonic radiation. This is also corroborated by our SBE results and is in agreement with recent studies concluding that above-bandgap harmonics are dominated by the interband recollision mechanism \cite{Vampa_2014, Yue_2020}. Moreover, since the harmonics studied here are, on average, \SI{3}{\electronvolt} above the bandgap, a two-band description—considering only the valence and conduction bands—is adequate. This is justified by the fact that, in our simulations, the combined energy width of these two bands exceeds \SI{8.6}{\electronvolt} (details in SI).

The phase variation of the XUV harmonics can be attributed to a modification of the high harmonic phase of the second XUV pulse. This modification arises from the interaction of the first NIR pulse with the sample, which alters the optical properties of the material.
The reasoning behind a positive or negative phase shift can be at least partially understood as follows: According to Eq.~\ref{eq:Pieters_expression} the phase decreases if the bandgap increases (note the minus sign in Eq.~\ref{eq:Pieters_expression}. The phase can be viewed as the accumulated action of the electron following excitation, hence a change in energy directly affects the phase. While the time the electron spends during acceleration also changes with changing bandgap, we found the influence of this change to be negligible compared to the band energy changes in our simulations. In contrast, if the bandgap decreases, the opposite occurs and the phase increases. \\
Several recent studies have proposed that in strongly excited fused silica, the bandgap tends to shrink \cite{Tsaturyan_2022, Tsaturyan_2024}. Although directly measuring the transient bandgap modulation is challenging - especially at intrapulse timescales \cite{Juergens_2024_LPR} - several indirect evidences have been obtained, indicating a significant shrinkage of the bandgap in photoexcited SiO$_2$\cite{Winkler_2020}, up to several eV. Conversely, in crystalline materials, an opposite behavior has been reported \cite{Feneberg_2014}. Due to Pauli blocking of the available and allowed energy states close to the conduction band edge and ground-state bleaching (i.e. a significant reduction of the ground-state population), the effective energy required for an electron to be promoted from the valence to the conduction band increases. This state blocking, or bandgap widening, has been observed in many different crystalline materials \cite{Saw_2015} and is often connected to the Buhrstein-Moss effect \cite{Burstein_1954}.
One possible explanation for the difference between amorphous and crystalline cases might be the lack of a well-defined band structure and energy bands in amorphous materials. In these materials, due to the existing crystalline short-range order, energy landscapes form in small clusters of the material. However, over a longer range or larger volume, the band structure washes out, resulting in a large number of available excited states. This inhibits the state blocking observed in crystalline materials, where only specific, well-defined energy levels are available for electron excitation. The microscopic mechanism leading to the bandgap narrowing in fused silica is still a topic of intense debate. Recent calculations suggest that strong-field-induced electronic charge redistribution may lead to a rearrangement of atoms, resulting in altered bonding strength and a corresponding decrease in the bandgap \cite{Tsaturyan_2022}. Additionally, intrinsic and photo-induced impurities and defects can modify the electronic landscape, thereby influencing the band structure and bandgap. 
Currently, no unified theory exists that simultaneously captures these effects alongside reliable modeling of photo-induced band structure variations, particularly for amorphous SiO$_2$. Consequently, calculations fully explaining the opposing trends observed for MgO and SiO$_2$ are beyond the scope of this work. However, our interpretation is strongly supported by previous experimental and theoretical studies. \\
 As described above, the opposite signs of the phase variation can approximately be assigned to different signs of the bandgap variation due to the presence of an electron-hole plasma. In Fig.~\ref{fig:final_figure}(a) and (b) we directly link $\overline{\Delta \phi}_{\rm{XUV}}$ to $\Delta E_g$ with the help of Eq.~\ref{eq:Pieters_expression}, which results in a maximum bandgap widening of $\sim$\SI{1}{\electronvolt} in the case of MgO [see Fig.~\ref{fig:final_figure}(b)] and a bandgap shrinkage of $\sim$\SI{0.8}{\electronvolt} for SiO$_2$ [Fig.~\ref{fig:final_figure}(a)] using characteristic excursion times of $\Delta t \sim \,$\SI{1.5}{\femto\second} as determined earlier.
 Such substantial bandgap variations are consistent with previous observations \cite{Winkler_2020} and numerical simulations of strong-field-induced bandgap shifts in wide-bandgap materials and indicate the possibility of tracking bandgap dynamics with sub-fs temporal precision using XUV spectral interferometry in wide-bandgap solids.  \\
We acknowledge that certain effects not currently included in our numerical models - such as multi-electron interactions, Berry-phase contributions, higher-lying bands, and the imaginary components of the tunneling excitation phase - could, in principle, influence the extracted bandgap shift. Consequently, Eq.~\ref{eq:Pieters_expression} represents an approximation. Nevertheless, we find the bandgap extraction highly insightful, as it underscores the broad applicability of our experimental technique for determining a key material property, which is of significant interest both for fundamental research and potential technological applications. \\
While these effects are not explicitly included in our numerical model, we can qualitatively assess their potential influence on the extracted bandgap shift. Many-body interactions, particularly carrier-induced screening, are expected to play a significant role as the electron-hole density approaches the critical density, leading to a reduction in the effective bandgap. The excitation of higher-lying bands modifies the density of states and screening behavior, potentially influencing the spectral response. Berry-phase effects and associated anomalous velocities introduce corrections to carrier dynamics that may contribute to transient modifications of the band structure. Additionally, the imaginary component of the tunneling excitation phase affects transition probabilities and dephasing, subtly shaping the evolution of the electronic structure. While these contributions add complexity, they are not expected to qualitatively alter the observed trend but rather refine the quantitative interpretation of the extracted bandgap shift. \\
Finally, the extracted frequency shifts and fringe ratios of the individual harmonics can be analyzed and interpreted in terms of the generated conduction band electron densities. Assuming that the frequency shifts of the harmonics are linked to an excitation-induced blueshift, a considerably higher density of photoexcited electron-hole pairs, despite a very similar bandgap, was generated in MgO. Additionally, the lifetime of the electron-hole plasma in MgO was reported as being significantly longer ($\tau_L \approx \,$\SI{1.5}{\pico\second} \cite{Guizard_1995}) compared to SiO$_2$ ($\tau_L \approx\,$\SI{150}{\femto\second} \cite{Martin_1997}).
Further evidence for higher plasma densities in MgO arises from the fringe contrast in the interferometric measurements (see Fig.~\ref{fig:add_exps}(c) for SiO$_2$ and (d) for MgO). As shown in Ref.~\cite{deKeijzer_2024}, a faster dephasing, characterized by a decrease of the dephasing time $T_2$, leads to a suppression of the HHG yield, which can be approximately described by:
\begin{equation}
    \label{eq:dephasing}
    Y \propto e^{-\frac{\Delta t}{T_2}}  T_2 \left( 1 - e^{-\frac{\Delta t}{T_2}} \right) .
\end{equation}
A carrier population generated by the leading pulse will enhance the electron-electron scattering rate and thus the dephasing during the HHG induced by the trailing pulse due to the presence of the electron-hole pairs as has been measured in photon-echo experiments in the single-photon weak-field limit \cite{Becker_1988}. This increased dephasing will then reduce the harmonic yield generated by the trailing pulse \cite{Heide_2022}. This reduced yield of the second interference source will result in a reduced amplitude of the oscillatory component and in an intensity-dependent reduction of the fringe contrast as experimentally observed in Fig.~\ref{fig:add_exps}(c),(d), compared to a perfect intensity modulation in the interferogram of two equal HHG emissions.
\section{Conclusion}
In conclusion, we have expanded high harmonic spectroscopy of solid-state systems to XUV spectral interferometry, a technique previously limited to gas phase and molecular systems. We analyzed the intensity-dependent variation of the high-harmonic phase using XUV spectral interferometry and correlated our experimental results with the transient alteration of the electronic structure. We observed fundamentally different behavior in amorphous systems (represented by SiO$_2$) and crystalline samples (represented by MgO). The experimentally observed differences were related to the distinct nature of the photoinduced bandgap dynamics and the consequent dipole phase dynamics. Specifically, we extracted bandgap shrinkage in the amorphous case and bandgap widening due to state blocking in the crystalline case. \\
Our results highlight the versatility of high harmonic spectroscopy for investigating ultrafast carrier dynamics and their resulting effects in solids. We also demonstrate how XUV interferometry provides an all-optical technique for band structure tomography with the potential for sub-cycle accuracy. The presented strategy is not limited to the specific classes of materials studied here but represents a widely applicable experimental technique with potential applications in the ultrafast metrology of semiconductors, thin films, and two-dimensional materials. Furthermore, high-harmonic interferometry may enable the decoding of various ultrafast electronic and structural dynamics when performed in different wavelength and delay ranges. Notably, the interferometric approach could also provide a viable strategy for all-optical probing of electron-hole coherence times, thereby offering direct access to the dephasing time of photoexcited electron-hole pairs.

\section*{Funding}
P. J. and A. H. acknowledge funding from the German Research Foundation (DFG, project no. 533828881). \\
Funding by the German Research Foundation - SFB1477 "Light-Matter Interaction at Interfaces," Project No. 441234705, is gratefully acknowledged. \\
Part of this work was conducted at the Advanced Research Center for Nanolithography, a public-private partnership between the University of Amsterdam (UvA), Vrije Universiteit Amsterdam (VU), Rijksuniversiteit Groningen (RUG), The Netherlands Organization for Scientific Research (NWO), and the semiconductor equipment manufacturer ASML and was (partly) financed by ‘Toeslag voor Topconsortia voor Kennis en Innovatie (TKI)’ from the Dutch Ministry of Economic Affairs and Climate Policy. This manuscript is part of a project that has received funding from the European Research Council (ERC) under the European Union’s Horizon Europe research and innovation programme (grant agreement no. 101041819, ERC Starting Grant ANACONDA), which funded P.v.E. and partly funded P.M.K. The project is also part of the VIDI research programme HIMALAYA with project number VI.Vidi.223.133 financed by NWO, which funded B.d.K. and partly funded P.M.K.

\section*{Disclosures}
The authors declare no conflicts of interest.

\section*{Data Availability Statement}
 Data underlying the results presented in this paper are not publicly available at this time but may be obtained from the authors upon reasonable request.

\section*{Supplemental document}
See Supporting Information for supporting content. 

\bibliography{XUV_IF} 

\newpage
\begin{center}
    {\Large Supporting Information: Extreme Ultraviolet High-Harmonic Interferometry of Excitation-Induced Bandgap Dynamics in Solids}  
    \\[1ex]
\end{center}

\section*{Supporting Information}
\subsection{Determination of absolute zero delay / temporal overlap}
\begin{figure}[ht]
    \centering\includegraphics[width=0.7\linewidth]{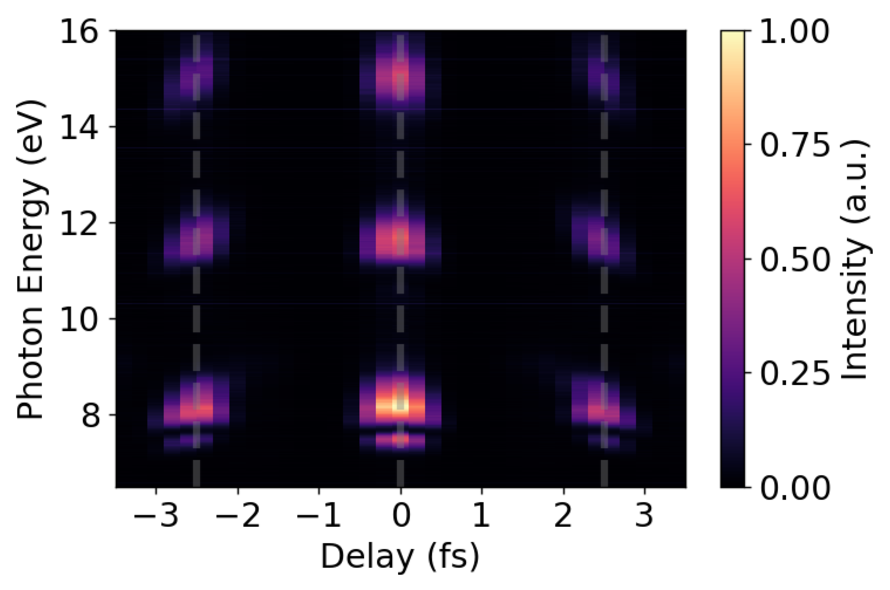}
    \caption{Experimental determination of the temporal overlap between the two phase-locked NIR copies. High-Harmonic spectrum generated in SiO$_2$ as a function of the relative delay between the two laser pulses. The vertical dashed lines indicate the position of an optical period at \SI{750}{\nano\meter}.}
    \label{fig:zero_delay}
\end{figure}

Using the actively stabilized Mach-Zehnder interferometer (MZI) shown in Fig.~1 of the main manuscript, we were able to perform highly accurate delay scans with sub-cycle precision. By analyzing the harmonic signal generated by the two few-cycle NIR pulses as a function of their relative delay, scanned in increments of \SI{0.2}{\femto\second}, significantly shorter than the optical period of the NIR field (i.e., \SI{2.5}{\femto\second}), we identified the absolute zero-delay position (see Fig.~\ref{fig:zero_delay}). This precise zero-delay reference enables accurate determination of the delay between the two phase-locked NIR pulses in subsequent experiments.

\subsection{Signal Optimization / Wedge Scan}
The few-cycle pulses were characterized using a SEA-F-SPIDER before being focused into the vacuum chamber for HHG. When harmonics are generated with an almost single-cycle pulse, the discrete harmonic spectrum broadens into a continuum-like XUV spectrum. Since this experiment aimed to investigate the variation of the interferometric fringe patterns of individual harmonics, we slightly detuned the dispersion-control wedges. This adjustment meant that harmonics were not necessarily generated with the shortest possible pulses from the solid targets. Nevertheless, the majority of the XUV signal (above-bandgap harmonics) was produced near the rear surface of the target. The additional dispersion introduced by the samples also required compensation. To address this, we performed wedge scans of the XUV interferometric spectra to identify optimal conditions. 
\subsection{Stability analysis}
\begin{figure}[ht]
    \centering\includegraphics[width=\linewidth]{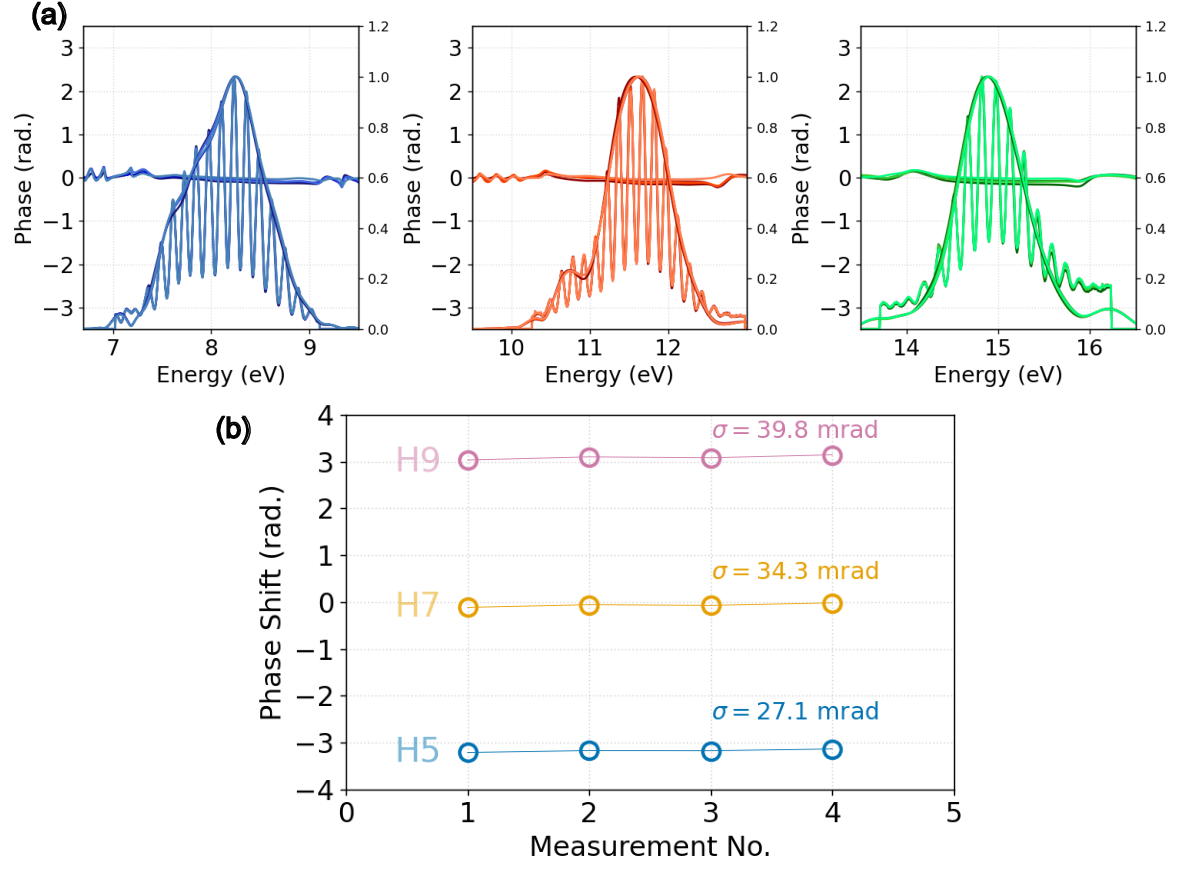}
    \caption{Control of the stability, reliability and reproducibility of the experimental procedure. (a) Extracted spectral phase of the harmonics obtained from four consecutive measurements under identical experimental conditions over a time period of \SI{30}{\minute} for the observed individual odd harmonics. (b) Weighted phase shift as a function of the measurement number. Phases for H5 and H9 are offset vertically for clarity.}
    \label{fig:stability}
\end{figure}

In attosecond and XUV interferometry measurements, it is essential to ensure that the stability of the experimental setup, along with the reliability and reproducibility of the data analysis, allows for the extraction of information with sufficient accuracy. For the setup used here, a thorough characterization was already performed in a prior study \cite{Koll_2022}, which quantified the stability to be better than \SI{10}{\atto\second}. In the present work, we re-evaluate the setups' capability to characterize sub-cycle phenomena by repeatedly measuring the interferometric XUV signal [similar to Fig.~2(a)] under identical conditions over a \SI{30}{\minute} period. Using the Takeda algorithm \cite{Takeda_1982}, we analyzed the phase variation during this period, with the first measurement serving as a reference. As shown in Fig.~\ref{fig:stability}, the phase variation was minimal, within the measurement uncertainty, with values ranging from \SI{-71}{\milli\radian} to \SI{6}{\milli\radian} for the fifth harmonic, \SI{-114}{\milli\radian} to \SI{-18}{\milli\radian} for the seventh harmonic, and \SI{-109}{\milli\radian} to \SI{2}{\milli\radian} for the ninth harmonic. Additionally, examining the fringe pattern, as obtained by integrating the harmonic signals along the divergence axis [see Fig.~\ref{fig:stability}(a)], shows no significant shifts in the positions of minima and maxima, thus confirming that the experimental apparatus is sufficiently stable to perform interferometric measurements with attosecond precision.

\begin{figure}[ht]
    \centering\includegraphics[width=\linewidth]{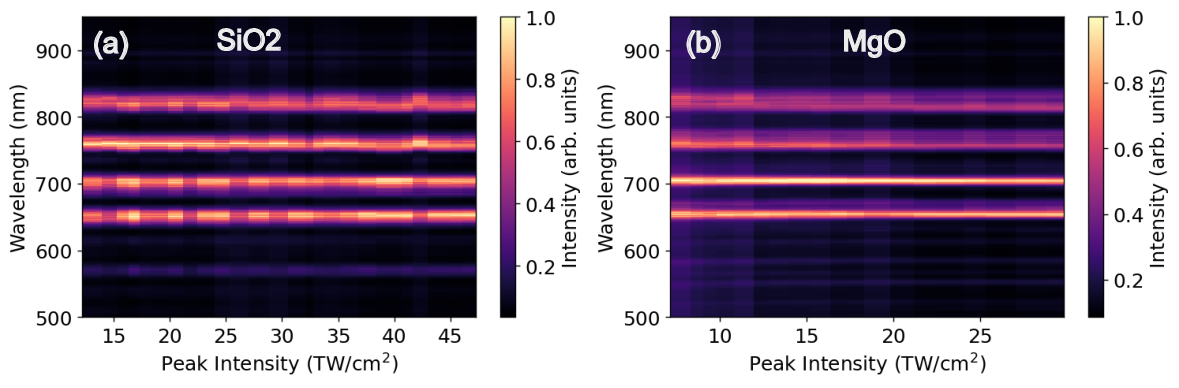}
    \caption{NIR Interferometry results in (a) SiO$_2$ and (b) MgO.}
    \label{fig:NIR_Maps}
\end{figure}

A comparable stability of the extracted NIR phase was demonstrated in Figs.~2 and 3 of the main manuscript. The interference maps shown in Fig.~\ref{fig:NIR_Maps} underpin the excellent stability of the MZI and the robustness of the NIR interferometry signal against variations in the NIR peak intensity.

\subsection{Numerical Estimate of NIR Phase Shift}
\begin{figure}[ht]
    \centering\includegraphics[width=0.75\linewidth]{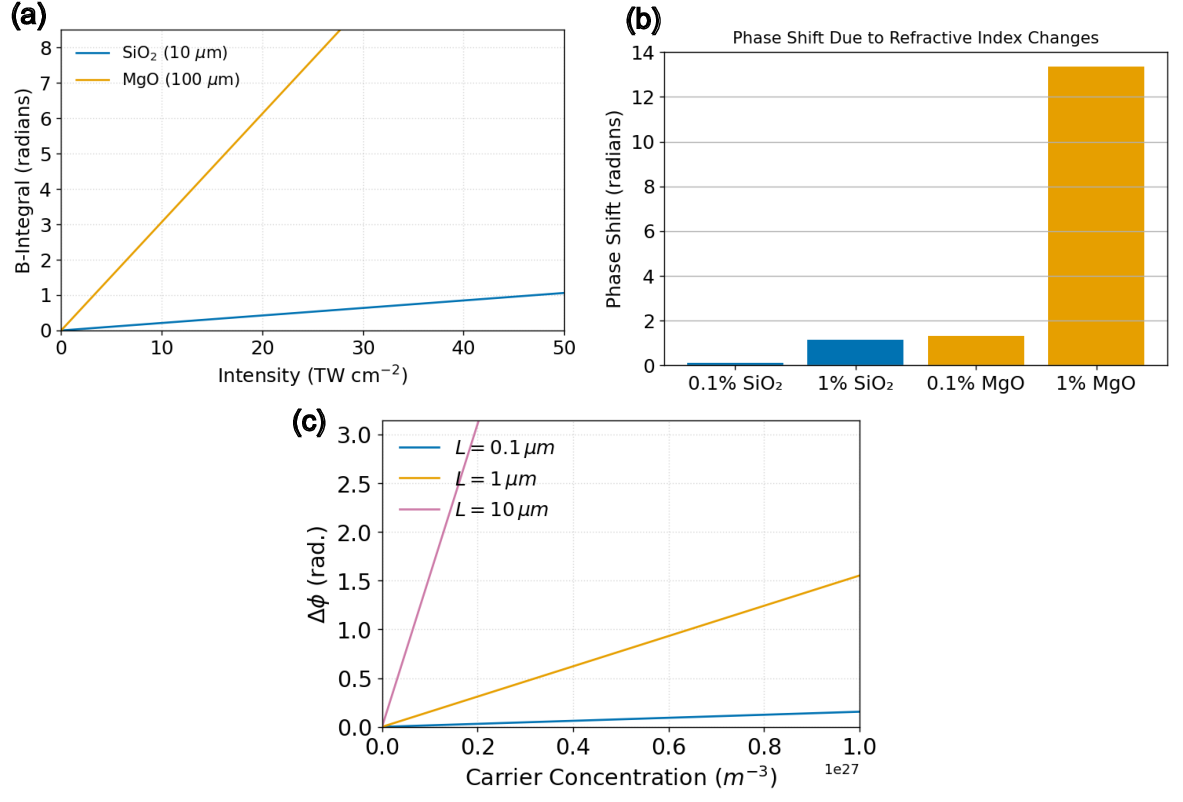}
    \caption{Numerical estimate of the NIR phase shift. (a) B-integral computed according to Equation~\ref{eq:B-Integral} for \SI{10}{\micro\meter} thick SiO$_2$ and \SI{100}{\micro\meter} thick MgO as a function of intensity. (b) Phase shift induced by a refractive index change of \SI{0.1}{\percent} (\SI{1}{\percent}) in SiO$_2$ and MgO. (c) Calculation of the NIR phase shift due to excitation-induced refractive index changes using the Drude model.}
    \label{fig:NIR_phase_estimate}
\end{figure}

A common parameter used to describe the accumulated nonlinear phase $\Delta \phi_{\rm{NL}}$ of an intense, ultrashort laser pulse propagating through a solid is the B-integral, defined as
\begin{equation}
    \label{eq:B-Integral}
    \Delta \phi_{\rm{NL}} = \frac{2 \pi}{\lambda} \int_{0}^{L} n_2 I(z) dz
\end{equation}
with central wavelength $\lambda$, nonlinear refractive index $n_2$ and laser intensity $I(z)$. In the case of loose focusing and thin samples the z-dependence of the laser intensity can be neglected, allowing us to assume a constant intensity throughout the entire sample thickness. A numerical estimate of the accumulated nonlinear phase as a function of laser intensity for SiO$_2$ ($n_2 =\,$\SI{2.7 e-20}{\meter\squared\per\watt}) and MgO ($n_2 = \,$\SI{3.9 e-20}{\meter\squared\per\watt}) is presented in Fig.~\ref{fig:NIR_phase_estimate}(a). While the accumulated phase remains relatively moderate in the case of the \SI{10}{\micro\meter} thick SiO$_2$ samples, it quickly exceeds a $2 \pi$ phase shift in the case of the \SI{100}{\micro\meter} thick MgO samples. Similarly, the phase shift $\Delta \phi = \frac{2 \pi}{\lambda} \Delta n L$ induced by refractive index changes of $0.1$ and \SI{1}{\percent}, that is summarized in Fig.~\ref{fig:NIR_phase_estimate}(b), shows the strong influence of $L$ for the MgO samples.   \\

It is important to note that these estimates overestimate the accumulated nonlinear phase, as they assume a constant peak intensity throughout the full thickness of the samples. In reality, particularly for few-cycle laser pulses, material dispersion significantly reshapes the pulse during propagation, limiting the region where the pulse maintains its shortest duration. In our experiments, positively chirped pulses impinge on the samples, and due to the normal dispersion of both SiO$_2$ and MgO, pulse compression occurs during propagation. As a result, the peak intensity is not reached at the front surface but only after a certain propagation distance. Consequently, assuming a constant peak intensity throughout the entire sample thickness leads to an overestimation of the accumulated nonlinear phase.

Finally, we aim at estimating the refractive index change and the associated phase shift using the Drude model, which describes the dielectric function of a free electron gas. The refractive index $n$ is related to the dielectric constant $\varepsilon$ by $n=\sqrt{\varepsilon}$. For an electron-hole plasma, the dielectric constant is given by:
\begin{equation}
    \varepsilon = \varepsilon_0 - \frac{\omega_{p}^{2}}{\omega^2}
\end{equation}
where, $\varepsilon_0$ is the dielectric constant of the unperturbed material, $\omega_p$ is the plasma frequency and $\omega$ is the angular frequency of the incident light. The plasma frequency is defined as:
\begin{equation}
    \omega_p = \sqrt{\frac{\rho e^2}{\varepsilon_0 m^{\ast}}}
\end{equation}
where $\rho$ denotes the carrier concentration, $e$ is the elementary charge and $m^{\ast}$ is the effective mass of the electrons. The phase shift $\Delta\phi$ experienced by a light wave traveling through a medium of thickness $L$ with a refractive index variation of $\Delta n$ is given by:
\begin{equation}
    \Delta \phi = \frac{2 \pi}{\lambda} \Delta n \times L
\end{equation}
where $\lambda$ is the wavelength of the NIR light in vacuum. 
A numerical estimate of the accumulated NIR phase for three different interaction lengths is shown in Fig.~\ref{fig:NIR_phase_estimate}(c). Clearly, besides the carrier density, the interaction length (i.e. the length over which a constant carrier concentration is present in the sample) determines whether a substantial phase shift is accumulated by the NIR field or not. \\
In summary, the absence of a clear phase shift in the NIR interferometry measurements can be attributed to the use of thin samples and short pulses, both of which effectively reduce the interaction length $L$ and thereby significantly limit the accumulated nonlinear phase.
Recollision-based high-harmonic signals are only emitted from the few last layers of the solid targets. We can estimate the thickness of the emitting layer by the penetration depth of the corresponding harmonic frequencies which is on the order of a few nm for the observed harmonics in the XUV spectral region. 
\subsection{Extraction of Fringe Depth}
To obtain the fringe contrast from the measurements, the experimentally observed high-harmonic spectra are projected onto the energy axis. All maxima and minima (further referred to as peaks and valleys) in the spectral interference pattern are identified with the help of a numerical routine. Then the ratio is calculated by dividing the spectral intensity in a maximum by the mean value of the spectral intensity of the two surrounding minima. To avoid complications due to a different number of identified peaks and valleys at different intensities the mean value of the ratios is taken at every overall NIR laser intensity. The ratios are computed for all laser intensities and all measurement runs and all observed harmonic orders.
\subsection{Detailed description of numerical models}
\subsubsection{Density Functional Theory (DFT)}
All calculations were performed using VASP 6.4.3\cite{VASP1,VASP2,VASP3,VASP4} with the HSE06 hybrid functional\cite{HSE06_original,HSE06_erratum}. The band structure calculations were performed on the bulk primitive cell and the MgO(100) surface cell comprising 11 crystallographic layers. To ensure Koopmans compliance, linearity corrections were applied following the methodology and parameterization of Wing \textit{et al.}\cite{wing2020}. The conduction band occupations were incrementally increased from \(10^{17}\) to \(10^{23} \, \text{cm}^{-3}\), and the electronic structure was relaxed. This approach isolates the effects of conduction band occupation on the band edges, ensuring any changes stem from occupancy rather than structural rearrangement. The band edge positions were referenced to the vacuum level, determined using the MgO(100) surface. Rather than focusing on the bulk contribution to the surface band structure described by Sagisaka and co-workers\cite{Sagisaka2017}, this approach was adapted to provide an absolute reference for the bulk band shifts. High-symmetry k-points for the band structure calculations were determined using the SeekPath utility\cite{HINUMA2017,togo2024}.
\subsubsection{Semiconductor Bloch Equations}
The simulation framework follows the methodology detailed in Ref.~\cite{deKeijzer_2024}, with additional specifics provided in \cite{Roscam_2024, Juergens_2024}. For completeness, a brief summary is presented here. To model HHG along a particular crystal axis, we utilize the semiconductor Bloch equations (SBEs) in the context of a one-dimensional two-band approximation. This framework effectively represents the valence and conduction bands of MgO along the $\Gamma$-X direction. The material parameters include a direct optical bandgap of $E_g =\,$\SI{7.8}{\electronvolt} \cite{Heo_2015}, with the valence and conduction band heights set to \SI{3.39}{\electronvolt} and \SI{5.25}{\electronvolt}, respectively. We did this to partly span the second-highest valence band. This yields enough absolute energy difference such that  H9 of 800 nm does not lie past the cut-off \cite{Vampa_2015_Semi}.

A lattice constant of $a =\,$\SI{4.19}{\angstrom} is used, and the transition dipole moment $d_k$ is estimated using first-order $\bf{k \cdot p}$ theory \cite{Haug2009,Golde2006}:
\begin{equation}
    d_k = d_0 \frac{E_g}{\epsilon_k^e - \epsilon_k^h}.
    \label{eq:dk}
\end{equation}
Here, $d_0 = 0.78$ a.u. represents the transition dipole moment at the $\Gamma$-point, and $\epsilon^{e(h)}_k$ denotes the single-particle energies of the conduction (valence) band. To calculate the dipole moment at the $\Gamma$-point, we utilize Quantum Espresso  with the generalized gradient approximation, specifically employing Perdew-Zunger functionals \cite{Giannozzi_2009}. Since the dipole moment is more sensitive to simulation parameters than to properties like energy bands, we confirmed the stability of our simulation results under small perturbations in $d_0$.

\begin{figure}[ht]
    \centering\includegraphics[width=\linewidth]{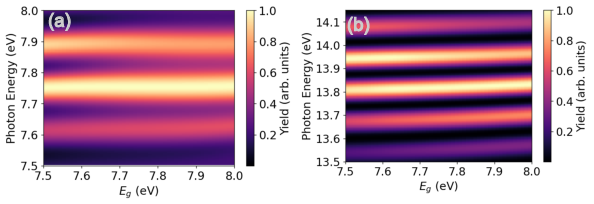}
    \caption{Additional SBE results for the spectral interferometry shifts of the fifth (a) and the ninth (b) harmonic in MgO.}
    \label{fig:supp_SBE}
\end{figure}

The simulations use a Gaussian laser pulse defined as:
\begin{equation}
    E(t) = E_0 \cos( \omega_0 t ) \exp\Big(-\Big( 2 \sqrt{\ln(2)} 
    \frac{t}{\tau} \Big)^2 \Big)
\end{equation}
where $E_0$ is the peak electric field, $\omega_0$ the central laser frequency, and $\tau$ the pulse duration. The semiconductor Bloch equations are solved while neglecting both carrier recombination and Coulomb interactions:
\begin{equation}
    i \hbar \frac{\partial}{\partial t} p_k = \Bigg(\epsilon_k^e + \epsilon_k^h - i \frac{\hbar}{T_2} \Bigg)p_k - (1- f_k^e-f_k^h) d_k E(t)+ieE(t) \cdot \nabla_k p_k,
 \label{eq:p2band}
\end{equation}
\begin{equation}
    \hbar \frac{\partial}{\partial t} f_k^{e(h)} = -2 \textrm{ Im}[d_kE(t)p_k^*] + eE(t) \cdot \nabla_k f_k^{e(h)}
     \label{eq:f2band}
\end{equation}
using sparse spectral methods \cite{dedalus} on a one-dimensional complex Fourier basis to compute the microscopic polarization $p_k$ and microscopic population $f^{e(h)}_k$. We used 129 points in $k$-space and a propagation time-step of \SI{10}{\atto\second} to reach convergence in our calculations.  \\
We calculate the HHG spectra from the macroscopic interband polarization
\begin{equation}
    P(t) = \sum_k[d_k p_k(t) + \mathrm{c.c.}]
\end{equation}
and the macroscopic intraband current
\begin{equation}
    J(t) = \sum_{\lambda,k} \mathrm{e} v_k^\lambda f_k^\lambda(t)
\end{equation}
with $\lambda \in \{e,h\}$ and group velocity
\begin{equation}
    v_g^\lambda = \frac{1}{\hbar} \frac{\text{d}}{\text{d}k}\epsilon_k^\lambda.
\end{equation}
By summing the individual contributions from the interband polarization and the intraband current we obtain the complex-valued amplitude of the spectrum emitted in the near-field as \cite{Schubert_2014}:
\begin{equation}
    A_\text{rad}(\omega) \propto  \omega^2 P(\omega) + i \omega J(\omega) .
    \label{eq:cplx_ampl}
\end{equation}
\begin{equation}
    I_\text{rad}(\omega) \propto |  A_\text{rad}(\omega) | ^2.
    \label{eq:near_field_full_weighted}
\end{equation}
To generate an interference pattern of the high-harmonic emission with transiently modified bandgaps, we calculate the complex-valued amplitudes (Eq.\ref{eq:cplx_ampl}) separately for a pristine bandgap and for shifted bandgap configurations. The emission amplitude for the shifted bandgap is modulated by a phase factor $\mathrm{e}^{-\mathrm{i} \omega \tau_{\rm{NIR-NIR}}}$, where $\tau_{\rm{NIR-NIR}}$ represents the delay between the two phase-locked laser pulses. The resulting complex-valued amplitudes are then added, and the interference pattern is determined using Eq.\ref{eq:near_field_full_weighted} from the combined complex amplitude.
SBE-results for showing interference maps of the fifth and the ninth order in MgO at a NIR peak intensity of \SI{30}{\tera\watt\per\centi\meter\squared} are displayed in Fig.~\ref{fig:supp_SBE} and exhibit a similar trend to the numerical results shown and discussed in the main manuscript. 
%
%
\subsubsection{Discussion and supporting analytical model}
Beyond the SBE calculations, we detail the analytical model employed in this study. In this approach, classical trajectories are calculated within a two-band system under the assumption of low carrier inversion, using the SBEs as a starting point. The semi-classical description is obtained by applying the interband saddle point approximation to the SBEs \cite{vampa_2017} and thus neglecting imperfect recollisions \cite{Yue_2020}. The analytical model employs the same band structure and transition dipole moment as those used in the SBE simulations.
Electron wavepacket trajectories are computed for a driving laser field within a single optical cycle. The external field dictates the effective carrier momentum for a given excitation time $t_i$ via,
\begin{equation}
    k(t) = A(t) - A(t_i) + k_0.
\end{equation}
with vector potential $A(t)$. Assuming a cosine driving field the vector potential reads:
\begin{equation}
    A(t) = -\int E(t) \text{ d}t = -\frac{E_0}{\omega_0}\text{sin}(\omega_0 t).
\end{equation}
Excitation is restricted to the $\Gamma$-point, where $k_0 = 0$, to ensure the unambiguous separation of long and short trajectories. Such distinct trajectories are not found using the SBEs, as these allow for excitation along the entire band structure \cite{Huttner_2016}. From a classical perspective, the results of the SBEs can be understood as a weighted average over all possible electron wavepacket trajectories. Consequently, a direct match between the SBE results and either the long or short trajectory results is not expected. Instead, the results from the long and short trajectories are anticipated to define upper and lower bounds for the SBE predictions. Electron wavepacket trajectories for different excitation times are determined through a purely classical integration of the group velocity:
\begin{equation}
    x^{\lambda}(t) = \int_{t_i}^{t}v^\lambda_g(k(\tau))\text{ d}\tau.
\end{equation}
Recombination is assumed to occur when the spatial displacement between the electron and hole reaches zero,
\begin{equation}
    \Delta x (t_r) = x^e(t_r) - x^h(t_r) = 0. 
\end{equation}
Here $t_r$ represents the recombination time of the trajectory, with only recombining trajectories contributing to the interband current. The photon energy emitted by a given trajectory corresponds to the energy difference between the charge carriers at the moment of recombination. The semi-classical model is limited to evaluating trajectories associated with harmonics above the band gap and below the cutoff.
To analyze the impact of band gap renormalization on the phase of the interband current, we evaluate the phase of the emitted light using the semi-classical action:
\begin{equation}
    S(t_r) = \int_{t_i}^{t_r} \Delta \epsilon(k(\tau)) \text{ d}\tau. 
\end{equation}
From the semi-classical action, the dipole phase is evaluated \cite{Carlstroem2016},
\begin{equation}
    \phi = \mathcal{N} \left(\omega_0 t_r + \frac{\pi}{2}\right) - S(t_r).
\end{equation}
Here, $\mathcal{N}$ denotes the harmonic order. 
Both the numerical and analytical models reveal a linear relationship between the phase and the bandgap variation. For small bandgap changes, the trajectories, along with their excitation and recombination times, are only minimally affected. Assuming moderate bandgap variations, the dipole phase can be approximated as follows:
\begin{equation}
    \Delta \phi \approx - \Delta E_g \Delta t \quad \text{with} \quad \Delta t = t_r-t_i.
\end{equation}
This rather simple linear relationship matches the numerical results of Ref.~\cite{deKeijzer_2024} and further indicates that $\Delta \phi$ depends on the $\Delta t$ of the different harmonic orders. As the SBEs do not include any classical trajectories, we should consider $\Delta t$ to be the characteristic excursion time of the harmonic rather than the trajectory time. 
From comparison of the semi-classical analytical model with the two-band SBE results we found that the effective excursion times are close to the average value of the long and short trajectory times which is  $\sim$\SI{1.5}{\femto\second} for the seventh harmonic order in the present case.


\end{document}